\documentclass[a4paper,11pt]{article}
\usepackage{jinstpub}
\usepackage{inputenc,graphicx}
\usepackage{siunitx}
\usepackage{multirow}
\usepackage{makecell}

\usepackage{lineno}
\newcommand{\vth}{$V_{\it{th}}$}
\newcommand{\sigmav}{$\sigma_{\it{V}}$}
\newcommand{\pdensity}{$P_{\it{density}}$}
\newcommand{\maxflu}{$1 \times 10^{16}$ n$_{\text{eq}}$/cm$^2$}
\newcommand{\flu}{n$_{\text{eq}}$/cm$^2$}

\title{Testbeam results of irradiated SiGe BiCMOS monolithic silicon pixel detector without internal gain layer}
\author[a]{T. Moretti,}
\author[a]{M. Milanesio,}
\author[a]{R. Cardella,}
\author[a]{T. Kugathasan,}
\author[a,b]{A. Picardi,}
\author[a]{I. Semendyaev,}
\author[c]{M. Elviretti,}
\author[c]{H. Rücker,}
\author[d]{K. Nakamura,}
\author[d]{Y. Takubo,}
\author[d]{M. Togawa,}
\author[a]{F. Cadoux,}
\author[a]{R. Cardarelli,}
\author[a]{L. Cecconi,}
\author[a]{S. Débieux,}
\author[a]{Y. Favre,}
\author[a]{C. A. Fenoglio,}
\author[a]{D. Ferrere,}
\author[a]{S. Gonzalez-Sevilla,}
\author[a]{L. Iodice,}
\author[a,b]{R. Kotitsa,}
\author[a]{C. Magliocca,}
\author[a,b]{M. Nessi,}
\author[a]{A. Pizarro-Medina,}
\author[a]{J. Sabater Iglesias,}
\author[a]{J. Saidi,}
\author[a]{M. Vicente Barreto Pinto,}
\author[a]{S. Zambito,}
\author[a,b]{L. Paolozzi,}
\author[a,1]{and G. Iacobucci\note{Corresponding author.}}

\affiliation[a]{Département de Physique Nucléaire et Corpusculaire (DPNC), University of Geneva, 24 Quai Ernest-Ansermet, CH-1211 Geneva 4, Switzerland}

\affiliation[b]{CERN, CH-1211 Geneva 23, Switzerland}

\affiliation[c]{IHP — Leibniz-Institut für innovative Mikroelektronik, Im Technologiepark 25, Frankfurt (Oder), Germany}

\affiliation[d]{High Energy Accelerator Research Organization, Oho 1-1, Tsukuba-shi, Ibaraki-ken, Japan}

\emailAdd{giuseppe.iacobucci@unige.ch}

\abstract{Samples of the monolithic silicon pixel ASIC prototype produced in 2022 within the framework of the Horizon 2020 MONOLITH ERC Advanced project were irradiated with 70 MeV protons up to a fluence of \maxflu, and then tested using a beam of 120 GeV/c pions. The ASIC contains a matrix of 100 $\mu$m pitch hexagonal pixels, read out by low noise and very fast frontend electronics produced in a 130 nm SiGe BiCMOS technology process. 
The dependence on the proton fluence of the efficiency and the time resolution of this prototype was measured
with the frontend electronics operated at a power density between 0.13 and 0.9 W/cm$^2$. 
The testbeam data show that the detection efficiency of 99.96\% measured at sensor bias voltage of 200 V before irradiation becomes 96.2\% after a fluence of \maxflu. 
An increase of the sensor bias voltage to 300 V provides an efficiency  to  99.7\% at that proton fluence. 
The timing resolution of 20 ps measured before irradiation rises for a proton fluence of \maxflu~to 53 and 45 ps at HV = 200 and 300 V, respectively. }

\begin{document}
\maketitle
\flushbottom
\section{Introduction}\label{sec:introduction}

To cope with the large event pile up foreseen during the CERN LHC High-Luminosity data-taking period, the experiments showed evidence for the need to upgrade the present detectors with layers with timing capability of the order of tens of picoseconds. 
The major LHC Collaborations foresee the addition of timing layers \cite{atlasTDR,cmsTDR} to match the required performance. 
One choice is to build these timing layers   with mm$^2$ silicon pads based on the low-gain avalanche detectors (LGAD)~\cite{PELLEGRINI201412}, which feature an internal gain layer under the pixel. 

The particle-physics community is presently attempting to develop a new  generation of silicon sensors able to achieve both high spatial granularity and excellent timing capabilities~\cite{Sadrozinski_2017,cartiglia}, although the radiation tolerance of the gain layer still constitutes a problem to place these sensors at small radii in present and future high-energy hadron colliders, where radiation will be very high. 
Studies to overcome the limited radiation tolerance of the gain layer are ongoing~\cite{SOLA2022167232,ASENOV2022167180,Croci_2023}.
A particularly interesting approach is to use a resistive layer that spreads the signal among four adjacent pixels and, by reconstructing the hit position from those signals, reduces drastically the number of detection channels needed to have spatial resolutions at the level of 10 $\mu$m, at the price of a reduction of the timing performance by approximately a factor of two~\cite{arcidiacono2022,KITA2023168009,imamura}.

The MONOLITH Horizon 2020 ERC Advanced project utilises the SG13G2 130 nm SiGe BiCMOS  process by IHP to produce low noise, low power and very fast frontend electronics,  implemented in a fully sensitive high granularity monolithic sensor able to provide excellent timing. 
The foundry masks of the first prototype of the MONOLITH project~\cite{Iacobucci_2022} were used to produce a proof-of-concept picosecond avalanche detector (PicoAD)~\cite{PicoADpatent}, a novel detector that implements a continuous deep gain layer~\cite{picoad_gain}. At a power density of 2.7 W/cm$^2$, this  proof-of-concept monolithic ASIC provided full efficiency and an average time resolution of 17 ps,  varying between 13 ps at the center of the pixel and 25 ps in the inter-pixel region~\cite{PicoAD_TB}.

\begin{figure}[!htb]
\centering
\includegraphics[width=.65\textwidth]{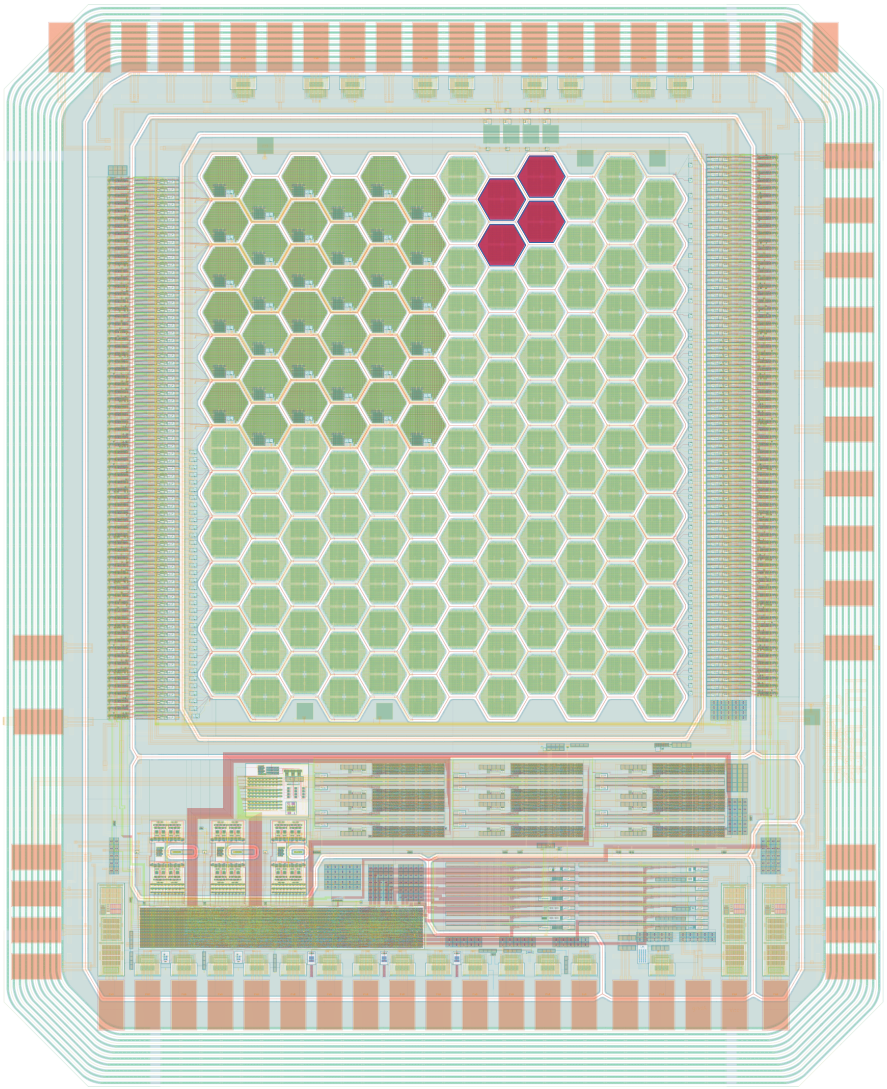}
\caption{\label{fig:layout} Layout view of the 2022 prototype ASIC. The analog pixels, in red, have the output of the amplifier directly connected to an analog driver with differential output.
}
\end{figure}

\begin{figure}[!htb]
\centering
\includegraphics[width=.70\textwidth]{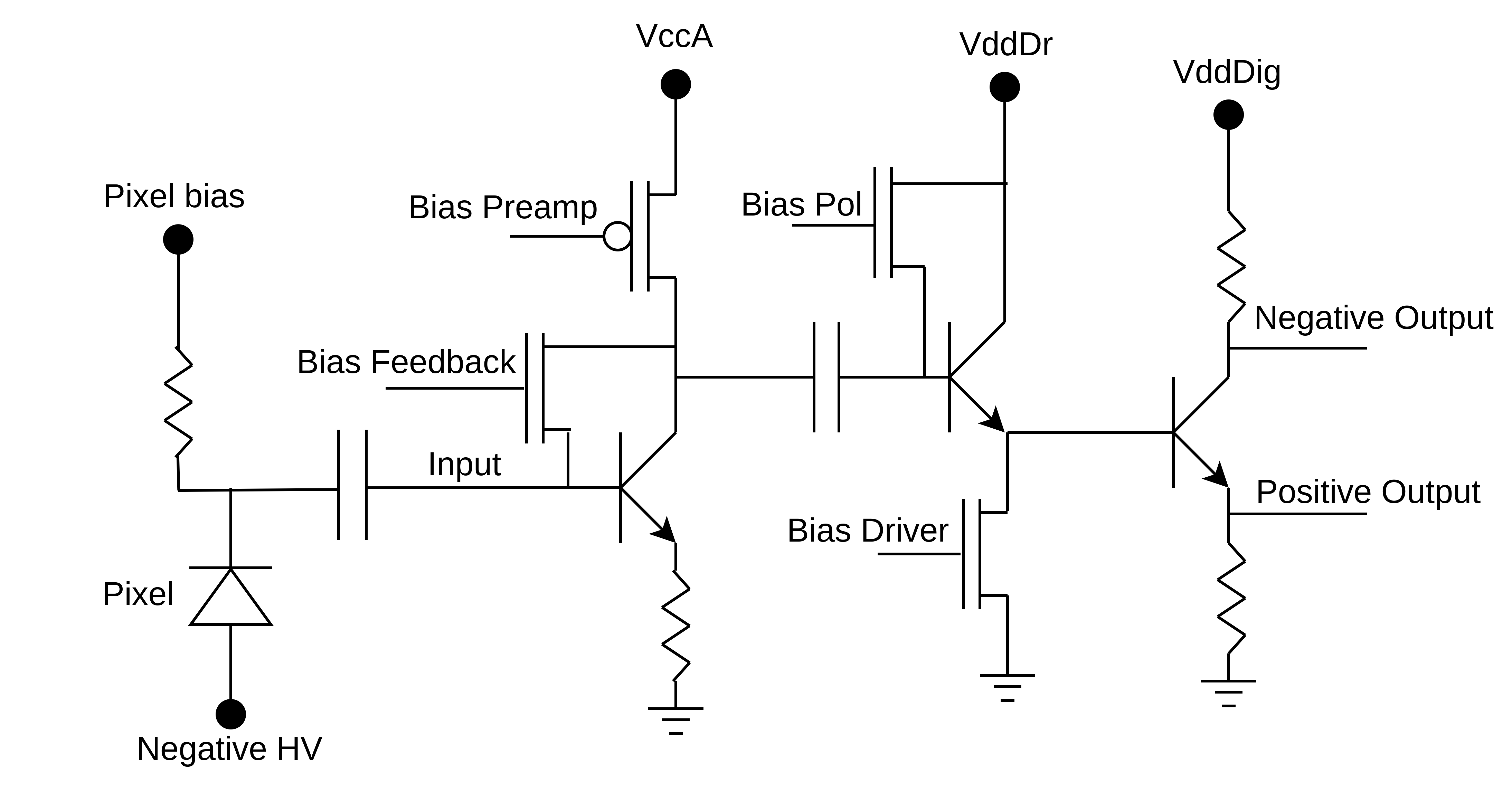}
\caption{\label{fig:frontend} Schematic view of the front-end and driver configuration of the analog pixels.
}
\end{figure}

A second prototype of the MONOLITH project containing an improved electronics~\cite{Zambito_2023} was produced in 2022. As for the first prototype, the new device contains four pixels where the amplifier was connected to an analog driver and could be read directly by an oscilloscope. Figure \ref{fig:layout} shows the layout of the 2022 prototype ASIC with the analog pixels highlighted in red. Figure \ref{fig:frontend} shows a schematic view of the front-end electronics and analog driver. 

Before the production of the next generation PicoAD prototypes using special wafers with deep gain-layer implant, the foundry masks of this second prototype were used to realise a detector without internal gain layer, implementing the sensor on a 50 $\mu$m thick epilayer of 350 $\Omega$cm resistivity.
The detection efficiency and time resolution of this prototype was measured~\cite{Zambito_2023} at the CERN SPS teastbeam facility with minimum ionising particles, and the time resolution as a function of the deposited charge was measured~\cite{Milanesio_2024} with a femtosecond laser.
%Indeed, recent prototypes in this framework have been characterized at the SPS testbeam facility at CERN with 120 Gev/c pions. The latest prototype, without internal gain layer, showed timing performances of 20 ps overall the full pixel thanks to improved frontend electronics and a higher resistivity of the substrate.
Several samples of this ASIC were irradiated at the CYRIC facility \cite{Nakamura_2015} in Japan with 70 MeV protons up to a fluence of \maxflu. 
%The pixel matrix contains four analog pixels consisting of a fast charge amplifier in SiGe HBT and a two-stages analog driver. 
The time jitter of the single-ended output of the  analog pixels was measured using a $^{90}$Sr radioactive source~\cite{milanesio2023radiation}.
In this paper, we present the efficiency and time resolution before and after proton irradiation measured with this prototype without gain layer using a beam of pions of 120 GeV/c at the CERN SPS.

%The following paper will present the results of the testbeam campaign of the same samples, this time at the SPS facility at CERN in which both the time resolution and the efficiency were measured for a Minimum Ionizing Particle (MIP).

\section{Data samples and experimental set-up}\label{sec:Samples&Setup}
Given the limited time availability during the testbeam experiment, data were taken only with four irradiated boards out of the seven characterised in \cite{milanesio2023radiation}. 
In addition, data were taken also with the board not irradiated previously characterised in \cite{Zambito_2023}.
Table \ref{tab:wp} gives a summary of the power density of the frontend and of the sensor bias voltage of the 18 datasets taken at the testbeam.
All boards were operated at a frontend feedback current of 2.0 $\mu$A and, to avoid any unwanted annealing,  stored and operated at a temperature of -10 $^\circ$C. 

\begin{table}[!htb]
\centering
\renewcommand{\arraystretch}{1.3}
\begin{tabular}{|c|ccc|cccc|}
\cline{1-8}
\begin{tabular}{c} Proton Fluence \\ \ [1 MeV n$_{\text{eq}}$/cm$^2$] \end{tabular} & \multicolumn{3}{c|}{\pdensity~ [W/cm$^2$]}  & \multicolumn{4}{c|}{High Voltage [V]} \\
\cline{1-8}
0                   & 0.13 & 0.54 & 0.9                                   & \multicolumn{4}{c|}{\multirow{4}{*}{200}} \\ \cline{1-4}
$9 \times 10^{13}$  & 0.13 & 0.54 & 0.9                                   & \multicolumn{4}{c|}{} \\ \cline{1-4}
$6 \times 10^{14}$  & 0.13 & 0.54 & 0.9                                   & \multicolumn{4}{c|}{} \\ \cline{1-4}
$3 \times 10^{15}$  & 0.13 & 0.54 & 0.9                                   & \multicolumn{4}{c|}{} \\ \hline
\multirow{2}{*}{$1 \times 10^{16}$ } & \multicolumn{3}{c|}{\begin{tabular}{cc}  0.13  & 0.54 \end{tabular}} & \multicolumn{4}{c|}{250} \\ \cline{2-8}
& \multicolumn{3}{c|}{0.9} & 150 & 200 & 250 & 300 \\
\cline{1-8} 
\end{tabular}
\caption{ Power density and sensor bias voltage of the 18 data samples taken at the CERN SPS testbeam with the four irradiated boards and with the board not irradiated.
The proton-fluence values  are reported in the first column in 1 MeV \flu~values.}
\label{tab:wp} 
\end{table}

The UNIGE FE-I4 telescope \cite{FEI4_telescope} was used to provide external tracking. 
The device under test (DUT)  was mounted at the center of the telescope
as explained in \cite{Zambito_2023}.
Two microchannel plate detectors (MCP) were positioned downstream of the telescope and used as time reference.
The analog outputs of the two MCP detectors  and of the DUT were read by two fast oscilloscopes. The first one with a sampling rate of 20 GS/s and an analog bandwidth set to 4 GHz for the channels connected to the MCPs and 1 GHz for the channels connected to the pair of single-ended analog output of the pixel used to measure the DUT time resolution. Two other pairs of single-ended output of adjacent pixels were connected to the second oscilloscope with a sampling rate of 10 GS/s and an analog bandwidth set to 1 GHz.
The DUT was mounted on a cooling plate and covered by an insulating box to keep the ASIC at a constant temperature of -10 $^\circ$C. 

The waveforms acquired for the positive and negative single-ended signals of the DUT were subtracted offline, sample point by sample point, to construct differential signals. Throughout the analysis, only the differential signals were used since they provide better signal-to-noise ratio.

\section{Detection efficiency}\label{sec:efficiency}

The detection efficiency was computed using pion tracks reconstructed by the FE-I4 telescope 
that produced hits in the six telescope planes and with a $\chi^2$/NDF $\le$ 1.5.
%and passing the track quality criteria consisting of described in \cite{Zambito_2023}. 
Since the number of channels available in the two oscilloscopes allowed recording of the signals from only two out of six neighboring analog pixels, the calculation of the efficiency was restricted to the telescope tracks whose extrapolation on the pixel surface was inside  the region defined by the triangle connecting the center of the three pixels (the pixel under study and the  two adjacent pixels that were read out). This triangle represents in the correct proportions all the areas of a pixel, and provide a result unbiased by the pointing resolution of the telescope which have been observed to show artificial inefficiency at the edge of pixels if the adjacent pixel is not read out~\cite{Zambito_2023}.

\begin{figure}[!htb]
\centering
\includegraphics[width=.85\textwidth]{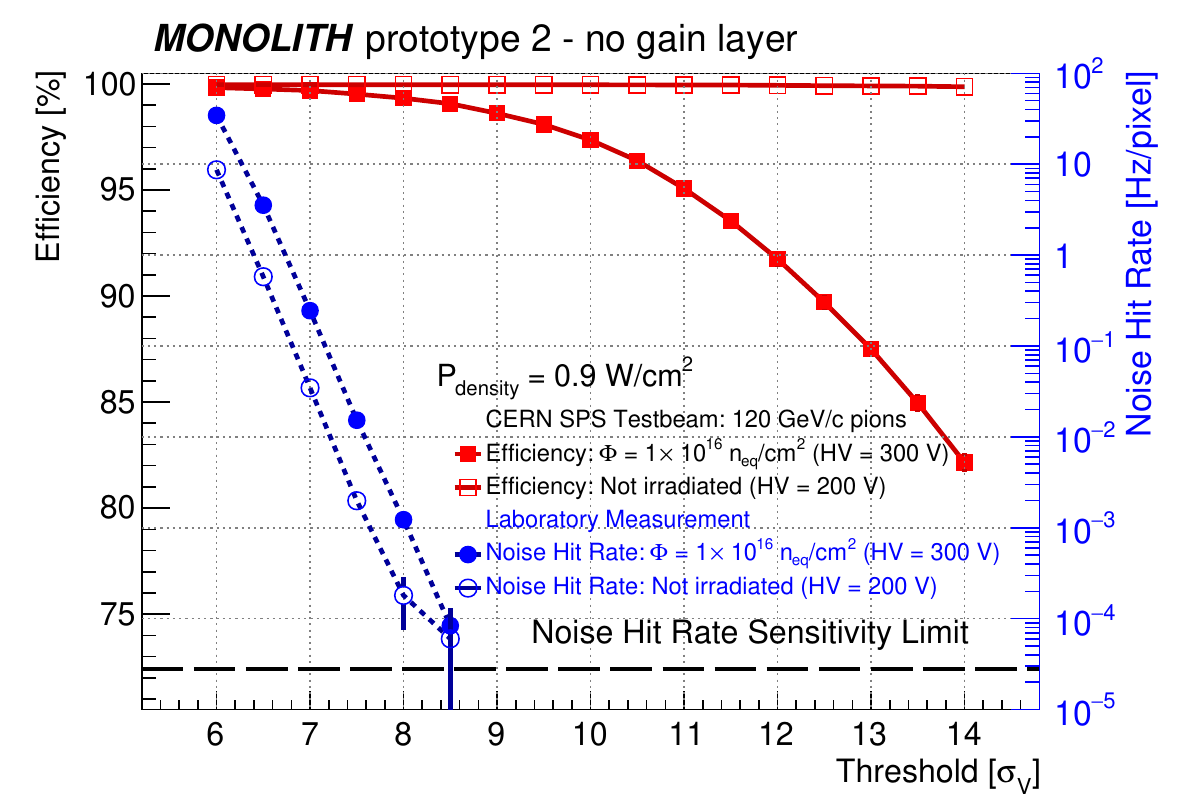}
\caption{\label{fig:HV_vs_th} Detection efficiency measured at a power density  \pdensity~= 0.9 W/cm$^2$ as a function of the voltage threshold. The data refer to the board not irradiated operated at HV = 200 V (open red squares), and to the board irradiated at \maxflu~operated at HV = 300 V (full red squares).
The voltage threshold is given in units of the voltage noise \sigmav, which is 300 $\mu$V before irradiation and raises to 600 $\mu$V after a fluence of \maxflu, as reported in~\cite{milanesio2023radiation}.
The plot reports also the noise-hit rate (values reported on the right-hand-axis scale) measured with the two boards.
}
\end{figure}
Figure~\ref{fig:HV_vs_th} shows the efficiency as a function of the voltage noise \sigmav, in the case of the board not irradiated operated at a sensor bias voltage of 200 V, and of the board irradiated at \maxflu~operated at 300 V. 
The increase of HV for the irradiated board was necessary to fully deplete the sensor and operate it at very high efficiency after the change in resistivity due to the exposure to radiation.
The voltage noise \sigmav~was measured event-by-event using the waveform samplings recorded in a time interval of 200 ns preceding the signal.
At the nominal threshold value \vth~= 7 \sigmav, the ASIC not irradiated provides a detection efficiency of (99.96$^{~\!+0.01}_{-0.02}$)\%
and, given the large signal-to-noise ratio, it maintains an efficiency at the level of 99.8\% even with a threshold value of 14 \sigmav.
On the contrary,  in the case  of the ASIC that was exposed to \maxflu, since the signal-to-noise ratio diminishes, the efficiency was found to depend more on the threshold, and at HV = 300 V varies from 99.7\% at a threshold \vth~= 7 \sigmav~to approximately 97.0\% already at \vth~= 10 \sigmav. 
Figure~\ref{fig:HV_vs_th} also reports the noise-hit rate as a function of the threshold. The noise-hit rate  shows the expected  exponential drop. It increases by a factor of approximately five after the ASIC has received a proton fluence of \maxflu.
\begin{figure}[!htb]
\centering
\includegraphics[width=.85\textwidth]{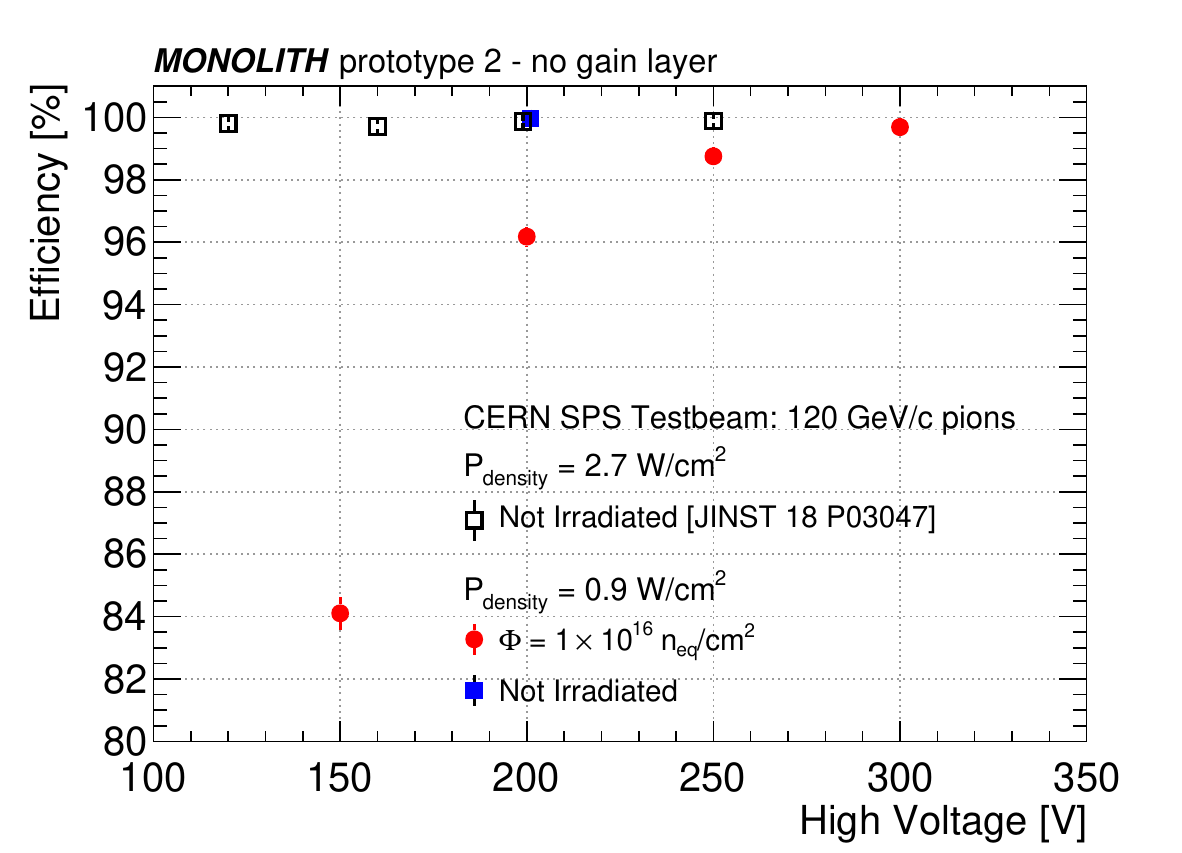}
\caption{\label{fig:HV_vs_eff} Detection efficiency measured as a function of the sensor bias voltage at a threshold value \vth~= 7 \sigmav. 
The red dots show the results obtained with the board irradiated at \maxflu, while the blue square that obtained with the board not irradiated, all  operated at \pdensity~= 0.9 W/cm$^2$.
The results obtained with the board not irradiated at \pdensity~= 2.7 W/cm$^2$ reported in \cite{Zambito_2023} are superimposed as black open squares.
}
\end{figure}

\begin{table}[!htb]
\centering
\renewcommand{\arraystretch}{1.3}
\begin{tabular}{|c|c|c|c|c|}
\cline{1-5}
%Proton Fluence & \pdensity & High Voltage &Detection & Time \\
%[1 MeV n$_{\text{eq}}$/cm$^2$] & [W/cm$^2$] & [V] &Efficiency Resolution \\
\cline{1-5}
\cline{1-5}
\begin{tabular}{c} Proton Fluence \\ \ [1 MeV n$_{\text{eq}}$/cm$^2$] \end{tabular} & \begin{tabular}{c}\pdensity~ \\ \ [W/cm$^2$] \end{tabular} & \begin{tabular}{c} High \\ Voltage \\ \ [V] \end{tabular} & \begin{tabular}{c}Detection \\ Efficiency \\ \ [\%]\end{tabular} & \begin{tabular}{c}Time \\ Resolution \\  \ [ps]\end{tabular} \\
\cline{1-5}
\multirow{3}{*}{ 0 }                    & 0.13 & \multirow{3}{*}{200}   & 99.94$^{~\!+0.03}_{-0.05}$ & 31.0 $\pm$ 1.6\\ \cline{2-2} \cline{4-5}
                                        & 0.54 &                        & 99.96$^{~\!+0.02}_{-0.03}$ & 29.0 $\pm$ 1.3\\ \cline{2-2} \cline{4-5}
                                        & 0.9  &                        & 99.96$^{~\!+0.01}_{-0.02}$ & 20.0 $\pm$ 1.0\\ \cline{1-5}
\multirow{3}{*}{ $9 \times 10^{13}$ }   & 0.13 & \multirow{3}{*}{200}   & 99.12$^{~\!+0.10}_{-0.19}$ & 37.8 $\pm$ 2.8\\ \cline{2-2} \cline{4-5}
                                        & 0.54 &                        & 99.63$^{~\!+0.07}_{-0.13}$ & 29.0 $\pm$ 1.3\\ \cline{2-2} \cline{4-5}
                                        & 0.9  &                        & 99.78$^{~\!+0.04}_{-0.11}$ & 24.7 $\pm$ 1.0\\ \cline{1-5} 
\multirow{3}{*}{ $6 \times 10^{14}$ }   & 0.13 & \multirow{3}{*}{200}   & 98.38$^{~\!+0.17}_{-0.19}$ & 42.4 $\pm$ 3.0\\ \cline{2-2} \cline{4-5}
                                        & 0.54 &                        & 99.13$^{~\!+0.12}_{-0.13}$ & 29.9 $\pm$ 2.2\\ \cline{2-2} \cline{4-5}
                                        & 0.9  &                        & 99.24$^{~\!+0.10}_{-0.11}$ & 26.4 $\pm$ 1.5\\ \cline{1-5} 
\multirow{3}{*}{ $3 \times 10^{15}$ }   & 0.13 & \multirow{3}{*}{200}   & 94.06$^{~\!+0.22}_{-0.23}$ & 75.9 $\pm$ 4.5\\ \cline{2-2} \cline{4-5}
                                        & 0.54 &                        & 97.22$^{~\!+0.17}_{-0.18}$ & 59.4 $\pm$ 0.9\\ \cline{2-2} \cline{4-5}
                                        & 0.9  &                        & 97.97$^{~\!+0.11}_{-0.11}$ & 48.3 $\pm$ 1.9\\ \cline{1-5} 
\multirow{6}{*}{ $1 \times 10^{16}$ }   & 0.13 & \multirow{2}{*}{250}   & 94.74$^{~\!+0.18}_{-0.19}$ & 74.0 $\pm$ 5.4\\ \cline{2-2} \cline{4-5}
                                        & 0.54 &                        & 98.33$^{~\!+0.10}_{-0.11}$ & 48.3 $\pm$ 1.9\\ \cline{2-5}
                    & \multirow{4}{*}{0.9}     & 150                    & 84.11$^{~\!+0.53}_{-0.54}$ & 60.4 $\pm$ 3.3\\ \cline{3-5} 
                                        &      & 200                    & 96.18$^{~\!+0.31}_{-0.33}$ & 53.1 $\pm$ 3.4\\ \cline{3-5}
                                        &      & 250                    & 98.75$^{~\!+0.09}_{-0.10}$ & 50.2 $\pm$ 1.6\\ \cline{3-5}
                                        &      & 300                    & 99.69$^{~\!+0.06}_{-0.07}$ & 45.3 $\pm$ 1.6\\ \cline{1-5}
% \cline{1-4}
\end{tabular}
\caption{Detection efficiency and time resolution measured for the 18 data samples taken at the CERN SPS testbeam with the four irradiated boards and with the board not irradiated.
The values were obtained for a signal amplitude threshold \vth~> 7~\sigmav.
}
\label{tab:effres} 
\end{table}

Table~\ref{tab:effres} reports the detection efficiencies obtained at a threshold \vth~= 7 \sigmav~for the 18 datasets listed in table~\ref{tab:wp}.

%The detection efficiency is expected to deteriorate as the signal-to-noise ratio becomes smaller from fluences of $3 \times 10^{15}$ \flu \cite{milanesio2023radiation} and above. To better understand the cause of this degradation it is important to study the performance while separating the contributions of the sensor from the front end electronics. 

Figure~\ref{fig:HV_vs_eff}  shows the detection efficiency of the ASIC irradiated at \maxflu~as a function of the sensor bias voltage. 
Only one data set at HV = 200 V was acquired at the testbeam with the board not irradiated. The data taken with the board not irradiated in a previous testbeam~\cite{Zambito_2023} are superimposed in figure~\ref{fig:HV_vs_eff} as open black squares for comparison\footnote{It should be noted that, in addition to the different power density, the data from \cite{Zambito_2023} were acquired at feedback current $i_{\it feedback}$ = 0.1 $\mu$A and at a temperature of 20$^\circ$C in contrast with the 2.0 $\mu$A and -10$^\circ$C of the present data. This change in operating conditions results in an increase of the signal-to-noise ratio, that leads to better efficiency and time resolution.}.
The measurement shows that after \maxflu~a bias voltage of 300 V is just enough to reach the detection efficiency plateau. This behaviour is completely different from that of the unirradiated board, for which the efficiency plateau is reached already at HV = 120 V.
%\footnote{At the time of the testbeam, it was not possible to operate the board irradiated with \maxflu~ at HV > 300 V without compromising the quality of the data taken, because anomalous spikes were triggering the data-acquisition system  preventing smooth data taking. This limitation in the operation of the irradiated sensors was attributed to radiation damage of the components on the boards that host the ASIC, in particular the HV decoupling capacitor. It was decided to delay the substitution of the capacitors to avoid the risk of damaging the boards before the testbeam. {\color{red} After the testbeam, the decoupling capacitor was substituted in one of the boards,  the rate of the spikes decreased by a factor of ten and the boards was successfully operated at larger HV values.}}.

This measurement, as well as the drop in detection efficiency shown in figure~\ref{fig:HV_vs_th}, could in large part be attributed to the change in resistivity of the silicon bulk, that we estimated to vary from the initial 350 $\Omega$cm to approximately 50 $\Omega$cm extrapolating the data of ~\cite{RadiationDamageBruzzi, DisplacementDamageMoll}.
At such resistivity, the 50 $\mu$m thick sensor would not be fully depleted.
A concurrent radiation damage of the frontend electronics cannot be excluded, and should be measured irradiating the SiGe HBT alone.

\begin{figure}[!htb]
\centering
\includegraphics[width=.85\textwidth]{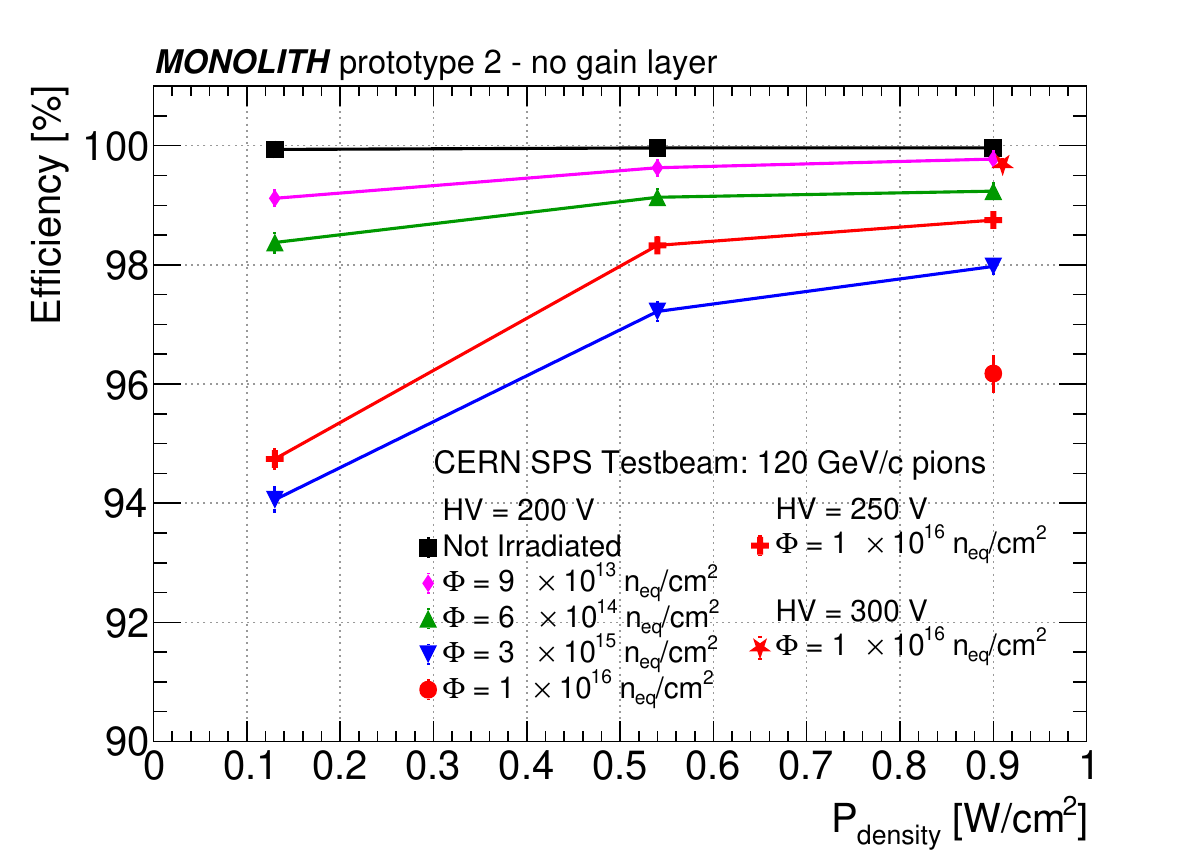}
\caption{\label{fig:Power_vs_eff}Detection efficiency measured for power density supplied to the preamplifier   between 0.13 and 0.9 W/cm$^2$ for a threshold value of \vth~= 7 \sigmav. The results obtained with the sample  not irradiated  (black empty squares) are presented with those of the four samples irradiated  with fluences between $9 \times 10^{13}$ \flu  and $1 \times 10^{16}$ \flu, all operated with a sensor bias voltage of HV = 200 V. 
In the case of the board irradiated at \maxflu~the data taken  at sensor bias voltage of 250 and 300 V are also shown.
The marker relative to the data point at HV = 300 V (red star) is displayed with a small horizontal offset to avoid  overlapping with other markers.}
\end{figure}

Figure~\ref{fig:Power_vs_eff} shows the impact on the detection efficiency of the power density at which the frontend was operated. 
At a sensor bias voltage HV= 200 V, a steady increase of the efficiency is measured as \pdensity~increases. This effect is more pronounced at the highest proton fluences. 

In the case of the board irradiated to \maxflu, at HV = 200 V data were taken only at \pdensity~= 0.9 W/cm$^2$, which result in a detection efficiency of (96.2 $\pm$ 0.3)\% (red circle in the figure).
For this board the scan in \pdensity~was performed at HV = 250 V, and the results are displayed in figure~\ref{fig:Power_vs_eff} by the red crosses. 
At HV = 300 V, only a dataset at \pdensity~= 0.9 W/cm$^2$ was acquired  (red star in the figure), which provides a detection efficiency of (99.69$^{~\!+0.06}_{-0.07}$)\%.

\begin{figure}[!htb]
\centering
\includegraphics[width=.85\textwidth]{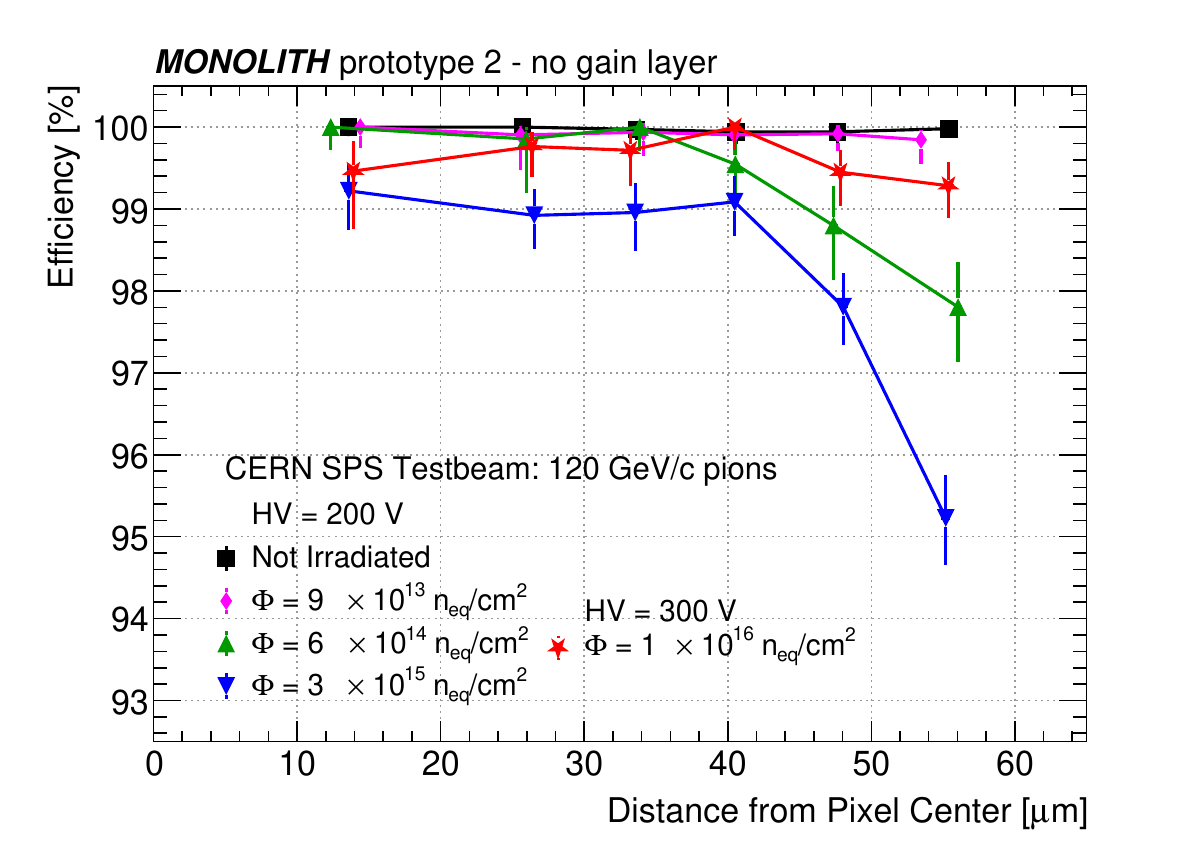}
\caption{\label{fig:Eff_vs_distance}Detection efficiency as a function of the distance from the pixel center. The data refer to power density \pdensity~= 0.9 W/cm$^2$, voltage threshold  \vth~= 7 \sigmav~and sensor bias voltage HV = 200 V, with the exception of the data taken at \maxflu, for which the sensor bias voltage was 300 V. The efficiency was computed in the following bins of distance from the pixel center in microns: $[0-21], [21-30], [30-37], [37-44], [44-51], [51-65]$. In each bin, the data points are plotted at the mean value of the track distance from the pixel center.}
\end{figure}

Finally, the detection efficiency is shown as a function of the distance from the centre of the pixel in figure~\ref{fig:Eff_vs_distance}.
All the data points shown refer to a sensor bias voltage of 200 V, with the exception of the dataset at \maxflu, which was obtained at 300 V.
While at no or small irradiation values the efficiency does not depend on the position within the pixel area, starting from a fluence of 6 $\times$ 10$^{14}$ \flu~the efficiency was measured to drop in the inter-pixel area, although an increase of the sensor bias voltage allows recovering of the efficiency far from the pixel center, as the data at HV = 300 V (red stars) show.

%\begin{figure}[!htb]
%\centering
%\includegraphics[width=.98\textwidth]{./Figures/Fluence_efficiency}
%\caption{\label{fig:fluence_vs_eff}Detection efficiency as a function of the fluence to which each sample was exposed, for a power density of $P_{density} = 0.9$ $W/cm^2$ and a discrimination threshold of $V_{th} = 7\cdot \sigma_V$. The results are presented for all the samples at a sensor bias voltage of HV = 200 V (blue circles). For the most irradiated sample, the detection efficiency is also shown for a sensor bias voltage of HV = 300 V (red star).}
%\end{figure}

%Finally, it is relevant to fix the sensor bias voltage and power density to their nominal values pre-irradiation of HV = 200 V and $P_{density} = 0.9$ $W/cm^2$ and see how the detection efficiency is affected. The results are presented in \ref{fig:fluence_vs_eff} and for the most irradiated sample, the value of the detection efficiency for a bias sensor voltage of HV = 300 V is also shown. The detection efficiency only drops by 2 \% after fluences of $3 \times 10^{15}$ \flu and by another 2 \% after $1 \times 10^{16}$ \flu. Arguably, full efficiency can easily be recovered by increasing the sensor bias voltage, as shown for the most irradiated sample.

\section{Time resolution}\label{sec:timing}

The time resolution of the DUT was measured using as reference the timestamp of two precise MCP detectors that were located outside the FE-I4 telescope downstream the pion beam. 
The sample of telescope tracks used for the measurement of the time resolution was the same used for the detection efficiency, with the only exception that all the tracks within the pixel were considered here, and not exclusively the tracks within the aforementioned triangle defined by the three analog pixel centers. 

The time of arrival (TOA) was measured for the DUT and the MCPs making offline a linear interpolation between the oscilloscope samplings and taking the time at which the signals reached 50\% of the maximum value of the signal amplitude.
The difference in TOA between the three pairs of detectors was then used to extract the time resolution of the DUT, as explained in~\cite{Zambito_2023}. 

Taking the TOA at a fixed percentage of the signal amplitude has the advantage to correct for most of the time walk between signals with different amplitudes, under the assumption that signals of all amplitudes have consistent rise times. 
It was observed that in our case this method carries the drawback of creating a small distortion of the distribution of the TOA difference, which deviates slightly from a Gaussian distribution and generates a tail at large values of ${\rm TOA}_{\it DUT}$ - ${\rm TOA}_{\it MCP}$,
probably from residual time walk not corrected for.
\begin{figure}[!htb]
\centering 
\includegraphics[width=.49\textwidth]{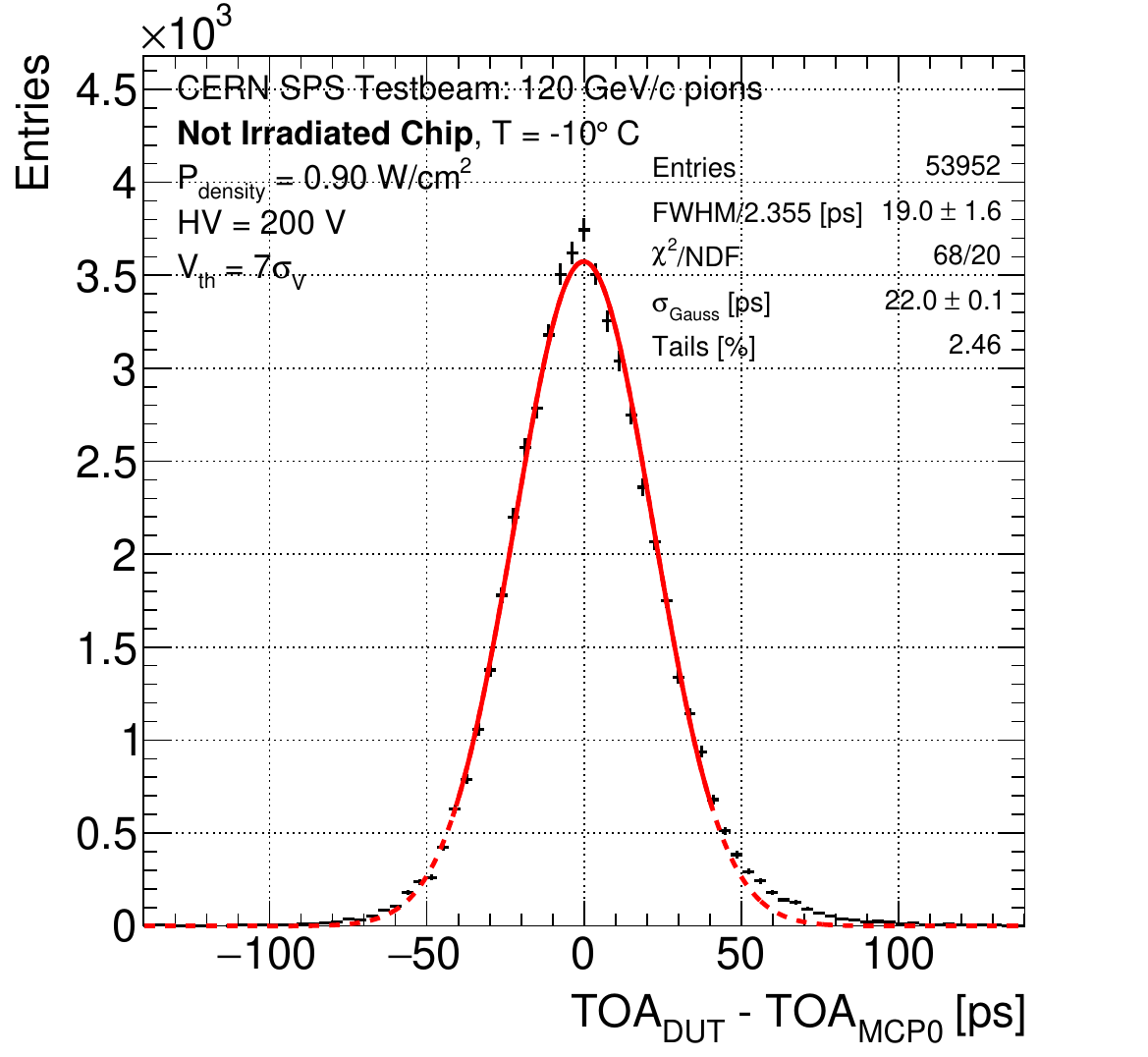}
\includegraphics[width=.49\textwidth]{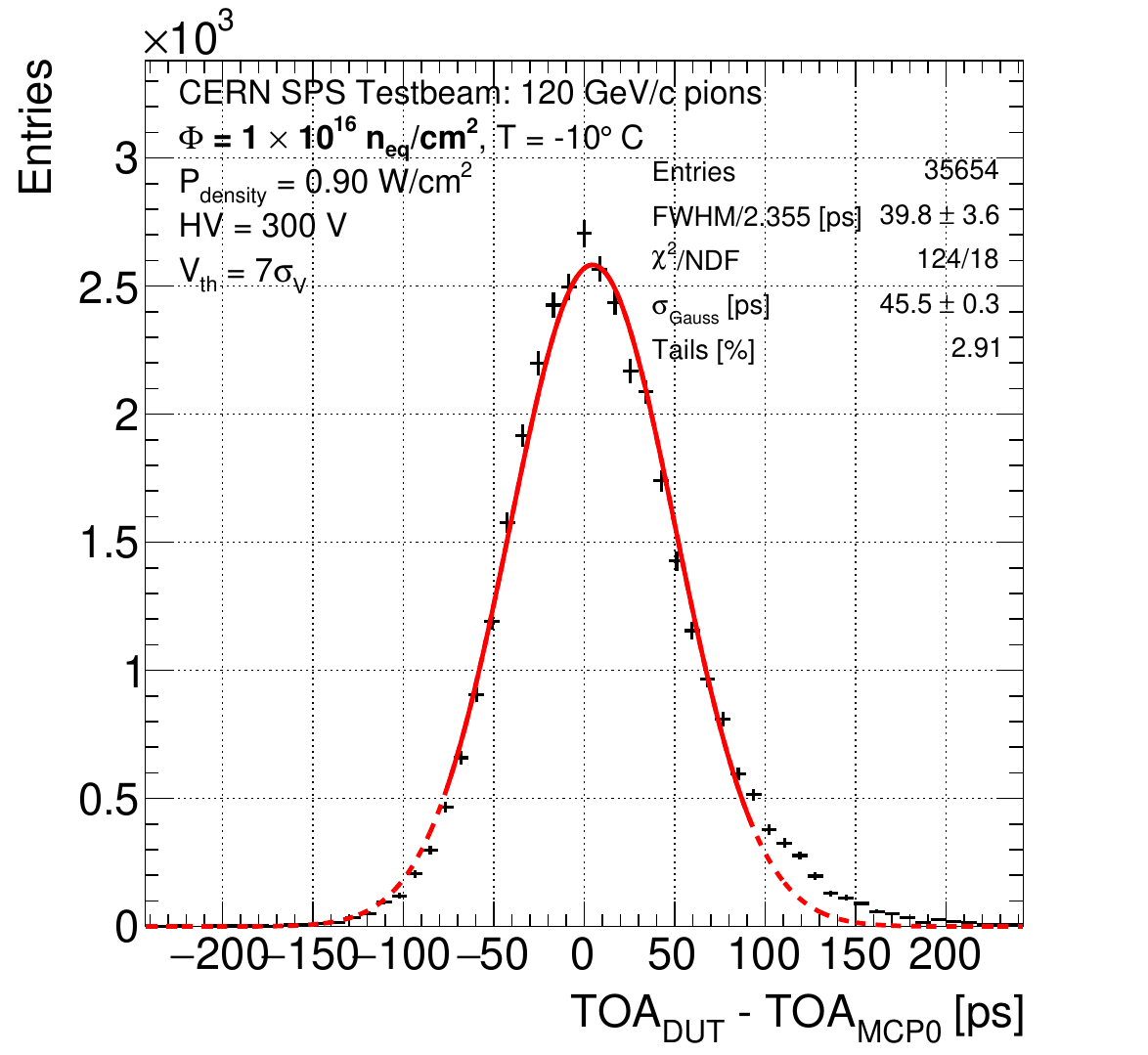}
\includegraphics[width=.49\textwidth]{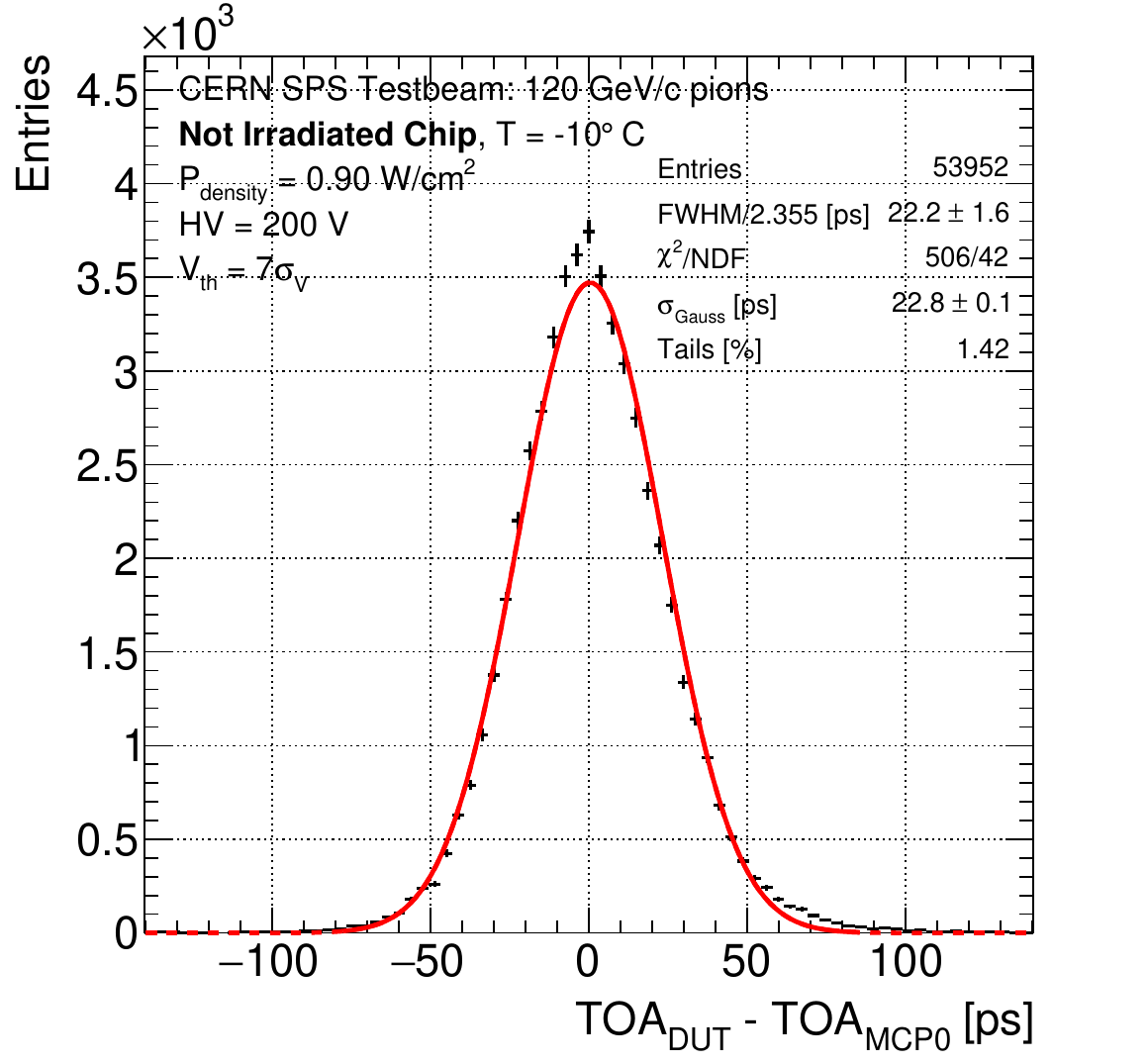}
\includegraphics[width=.49\textwidth]{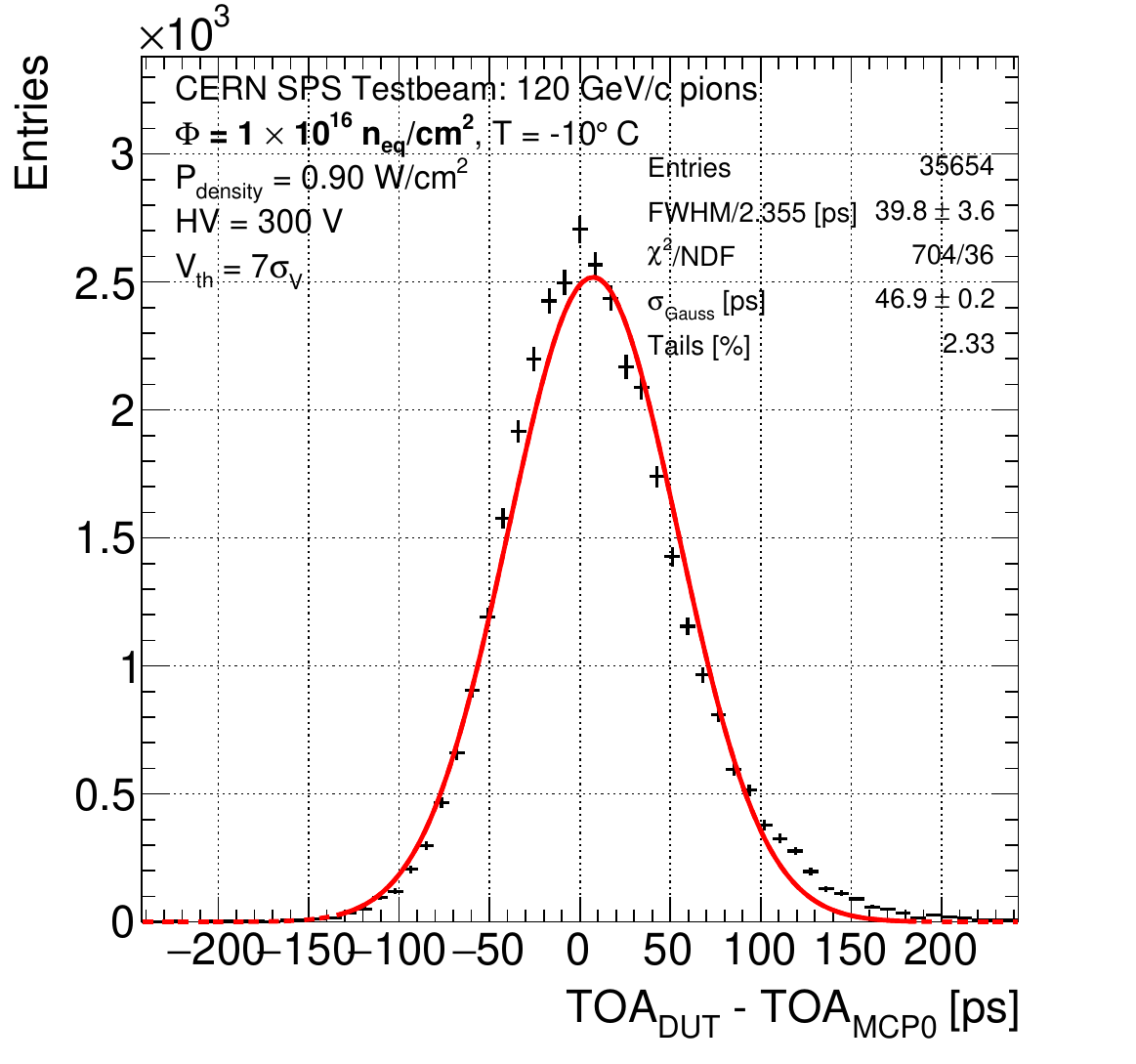}
\caption{\label{fig:toa} 
Difference in TOA between the DUT and one of the MCP detectors used to provide a precise reference time.
The left-hand panels show the results for the board not irradiated, while the right-hand panels those for the board irradiated to \maxflu.
The top panels show the data with superimposed the Gaussian fits extending to the first bin containing 20\% of the maximum value of the distribution; 
the bottom panels show the same data with superimposed the Gaussian fits extending to all the bins of the distributions.
%The $\sigma$ values resulting from the fits of the top panels are used as the central value of the time resolutions, while their difference w.r.t. the bottom plots are summed in quadrature to the statistical uncertainty on the $\sigma$ value from the fits and used to estimate the uncertainty of the time resolution measurements.
}
\end{figure}
This effect is shown in figure~\ref{fig:toa} in the case of the board not irradiated (left panels) and of the board irradiated at \maxflu~(right panels). 
%The deviation from a Gaussian distribution was found to increase for increasing irradiation, producing a tail at the level of 2\% in the case of the board not irradiated, that increased to 5\% in the case of the board irradiated at \maxflu.

To account for the observed distortion in the TOA difference, the $\sigma$ values from the Gaussian fits obtained using only the bins of the distribution larger than 20\% of the  bin with the highest number of entries (two top panels in the figure~\ref{fig:toa}) were used as central values of the  time resolution.
The difference between these values and those obtained by fitting a Gaussian functional form in the entire TOA-difference distributions (bottom panels in figure~\ref{fig:toa}) were used as systematic uncertainty. These uncertainties were summed in quadrature with the statistical uncertainty coming from the Gaussian fits that were used as central values, and are shown as vertical error bars in the  time resolution plots.

\begin{figure}[!htb]
\centering
\includegraphics[width=.85\textwidth]{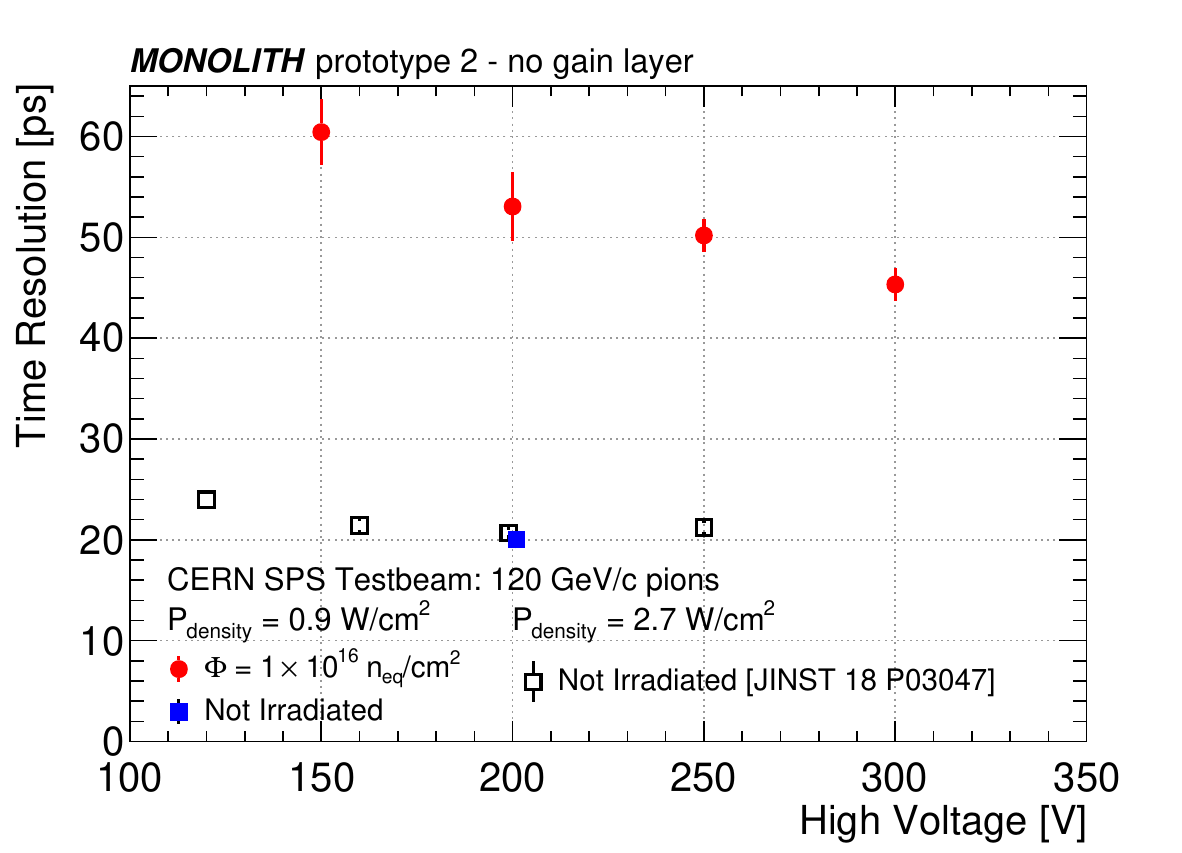}
\caption{\label{fig:HV_vs_timeres} Time resolution measured as a function of the sensor bias voltage. The results 
obtained with the board irradiated to \maxflu~at \pdensity~= 0.9 W/cm$^2$ (red circles) are presented together with the result obtained with the board not irradiated at HV = 200 V (blue square).
The plot results obtained in a previous testbeam~\cite{Zambito_2023} with the board not irradiated at \pdensity~= 2.7 W/cm$^2$ and  $i_{\it feedback}$ = 0.1 $\mu$A are also superimposed (black open squares).
}
\end{figure}

The last column of table~\ref{tab:effres} reports the time resolution obtained for the 18 datasets acquired at \vth~= 7 \sigmav.

Figure~\ref{fig:HV_vs_timeres} shows the time resolution obtained at \pdensity~= 0.9 W/cm$^2$ for the board irradiated to \maxflu~for a sensor bias voltage varying between 150 and 300 V, and the result obtained with the unirradiated board at HV = 200 V.
For comparison, the figure reports as well the time resolution measured at \pdensity~= 2.7 W/cm$^2$ with the board not irradiated in  previous testbeam experiment~\cite{Zambito_2023}.
While the board not irradiated reaches the plateau of time resolution already at HV = 160 V, the time resolution measured with the irradiated board does not seem to have yet reached a plateau value at 300 V.

\begin{figure}[!htb]
\centering
\includegraphics[width=.85\textwidth]{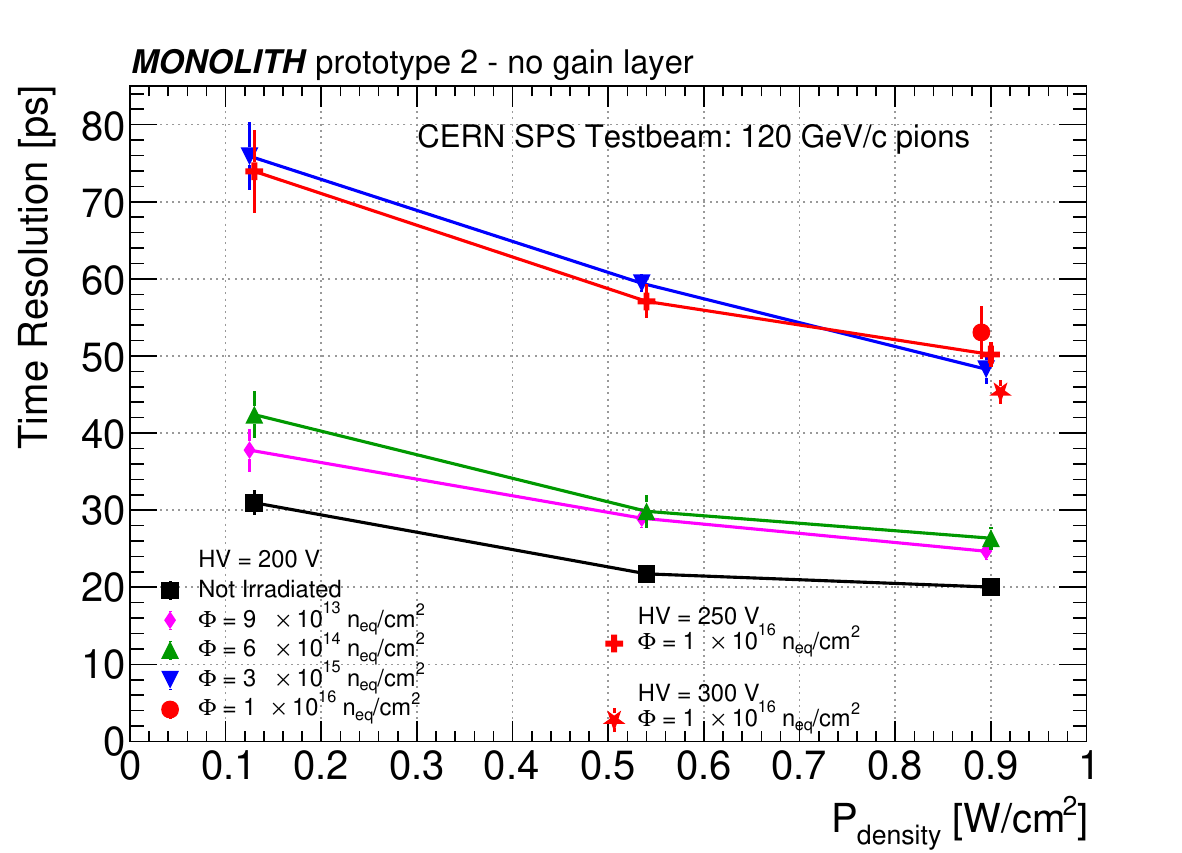}
\caption{\label{fig:power_vs_timeres}Time resolution measured for three values of the  power density supplied to the preamplifier between 0.13 and 0.9 W/cm$^2$. The plot shows the results obtained at sensor bias voltage HV = 200 V with the five boards which were subject to proton fluence between zero and \maxflu. 
The data taken with the board most irradiated  at HV = 250 and 300 V are also shown.
}
\end{figure}

The time resolution of the DUT was also measured as a function of the power density supplied to the frontend electronics. The results are presented in figure~\ref{fig:power_vs_timeres} at a sensor bias voltage HV = 200 V  for all the boards, and for the most irradiated board also at HV = 250 and 300 V. 
In all cases, the time resolution was found to steadily improve with increasing \pdensity, and degrade with increasing proton fluence. In the \pdensity~range studied, it varies between 30 and 20 ps in the case of the board not irradiated, and between 75 and  50 ps in the case of the board irradiated to \maxflu.

It should be noted that, for each of the irradiation values studied, the ratio of the time resolutions at \pdensity~= 0.13 and 0.9 W/cm$^2$ is constant, showing a relative improvement of approximately 35\% independently of the fluence, as reported in table~\ref{tab:Power_improvement}.  
%These results might indicate that even the very high irradiation level studied here {\color{green}had no effect on the performance of the frontend electronics, going in the same direction as what was observed in~\cite{milanesio2023radiation}.}  

\begin{table}[!htb]
\centering
\renewcommand{\arraystretch}{1.2}
\begin{tabular}{|c|c|c|c|}
\cline{1-4}
\multirow{2}{*}{\begin{tabular}{c}Fluence \\ \ [1 MeV n$_{\text{eq}}$/cm$^2$]\end{tabular} } & \multicolumn{2}{c|}{\begin{tabular}{c}  Time Resolution [ps] \end{tabular}} & \multirow{2}{*}{\begin{tabular}{c} Ratio  \end{tabular}} \\ \cline{2-3}
&\pdensity~ = 0.13 W/cm$^2$&\pdensity~ = 0.9 W/cm$^2$& \\
\cline{1-4}
0                   & 31.0 $\pm$ 1.6 & 20.0 $\pm$ 1.0 & 1.55 $\pm$ 0.15 \\ \hline
$9 \times 10^{13}$  & 37.8 $\pm$ 2.8 & 24.7 $\pm$ 1.0 & 1.53 $\pm$ 0.21 \\ \hline
$6 \times 10^{14}$  & 42.4 $\pm$ 3.0 & 26.4 $\pm$ 1.5 & 1.61 $\pm$ 0.21 \\ \hline
$3 \times 10^{15}$  & 75.9 $\pm$ 4.5 & 48.3 $\pm$ 1.9 & 1.57 $\pm$ 0.17 \\ \hline
$1 \times 10^{16}$  & 74.0 $\pm$ 5.4 & 50.2 $\pm$ 1.6 & 1.57 $\pm$ 0.19 \\ \hline
\end{tabular}
\caption{\label{tab:Power_improvement} Time resolution measured with the five boards at \pdensity~= 0.13 and 0.9 W/cm$^2$ and their ratio. The data were taken at sensor bias voltage HV = 200 V for all boards except the board irradiated to \maxflu~for which the sensor bias voltage was 250 V.}
\end{table}

This is an indication that the dependence of the transition frequency from the collector current, which is observed in \cite{SiGeRadiationDamage} not to change up to 1 $\times$ 10$^{14}$ \flu, might not be affected by radiation damage even at \maxflu. 

\begin{figure}[!htb]
\centering
\includegraphics[width=.67\textwidth]{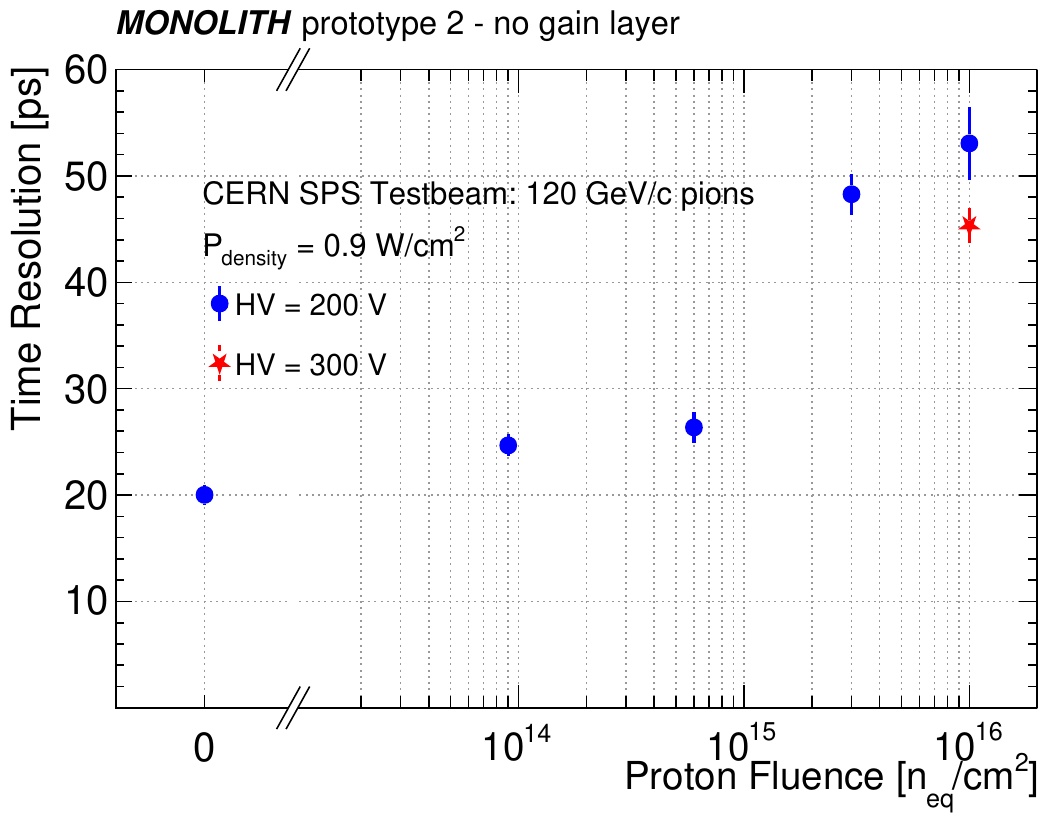}
\caption{\label{fig:fluence_vs_timeres} Time resolution as a function of the proton fluence for \pdensity~= 0.9 W/cm$^2$. The blue dots show the results obtained at a sensor bias voltage HV = 200 V, while the red star shows the result obtained at HV = 300 V with the board irradiated to \maxflu.
}
\end{figure}

Figure~\ref{fig:fluence_vs_timeres} shows the time resolutions as a function of the proton fluence at \pdensity~= 0.9 W/cm$^2$. The results obtained with the datasets taken at sensor bias voltage HV = 200 V vary between 20 ps for the board not irradiated and 53 ps for the board irradiated to \maxflu. 
For this last board, an increase of the sensor bias voltage to 300 V provides a time resolution of 45 ps, thus improving it by approximately 10\%.

\begin{figure}[!htb]
\centering
\includegraphics[width=.67\textwidth]{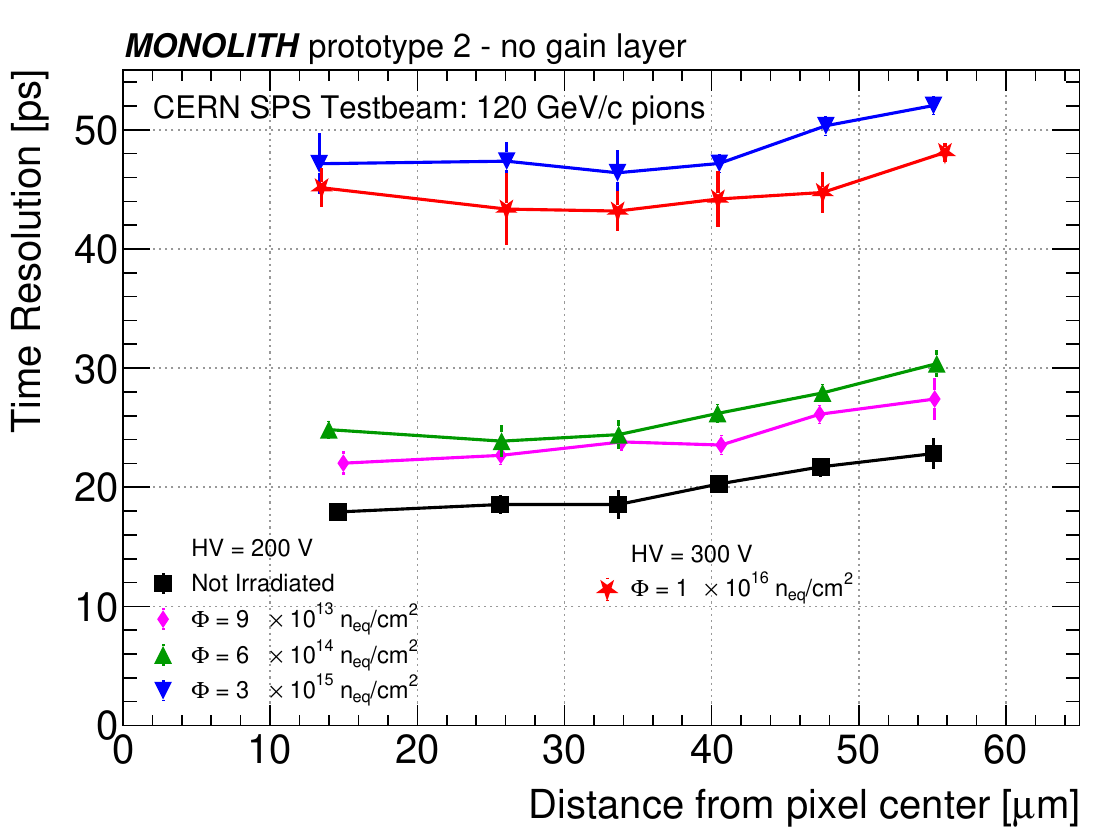}
\caption{\label{fig:Timeres_vs_distance}Time resolution as a function of the distance from the pixel center. The data refer to \pdensity~= 0.9 W/cm$^2$,  \vth~= 7 \sigmav~and HV = 200 V, with the exception of the data taken at \maxflu, for which the sensor bias voltage was 300 V. The time resolution was computed in the 
same bins of figure~\ref{fig:Eff_vs_distance}.
%following bins of distance from the pixel center in microns: $[0-21], [21-30], [30-37], [37-44], [44-51], [51-65]$. In each bin, the data points are plotted at the mean value of the track distance from the pixel center.
}
\end{figure}

The time resolution is shown as a function of the distance from the centre of the pixel in figure~\ref{fig:Timeres_vs_distance}.
All the data refer to a sensor bias voltage of 200 V, with the exception of the dataset at \maxflu, which was acquired at 300 V.
For all fluences the time resolution is better in the central region of the pixel than in the inter-pixel region, where it increases approximately 5 ps.

\section{Summary and Conclusions}\label{sec:conclusions}
A monolithic silicon pixel prototype produced for the Horizon 2020 MONOLITH ERC Advanced project was irradiated up to a proton fluence of \maxflu.
The prototype contains a matrix of hexagonal pixels with 100 $\mu$m pitch, read by a SiGe BiCMOS 130 nm frontend electronics. The sensor was produced on a 50 $\mu$m epilayer of a resistivity of 350 $\Omega$cm. 
The analog channels present in the pixel matrix were read by a fast oscilloscope and tested at the CERN SPS using a beam of 120 GeV/c pions. 
Only signals with an amplitude larger than a voltage threshold  \vth~= 7 \sigmav~were considered for the measurement of the detection efficiency and  time resolution.

Before proton irradiation, the detection efficiency at a sensor bias voltage HV = 200 V and frontend  power density \pdensity~= 0.9 W/cm$^2$ was measured to be (99.96$^{~\!+0.01}_{-0.02}$)\%.  For a proton fluence  \maxflu, the detection efficiency in the same conditions was measured to be (96.2 $\pm$ 0.3)\%; an increase of the bias voltage to 300 V provided (99.69$^{~\!+0.06}_{-0.07}$)\%.
%, thus approaching the voltage plateau of the sensor. 
These efficiency values include also the inter-pixel regions. Therefore there is no inactive area in between adjacent pixels within the sensor matrix.

The time of arrival of the signals used in the data analysis was defined as the time at 50\% of the signal amplitude, and was computed by a linear interpolation of the oscilloscope samplings. No further time-walk correction was applied to the data. 
The time resolution at  \pdensity~= 0.9 W/cm$^2$ and HV = 200 V was measured to be (20.0 $\pm$ 1.0) ps before proton irradiation. 
At the highest proton fluence studied of \maxflu, the time resolution became (53.1 $\pm$ 3.4) ps. 
An increase of the sensor bias voltage to HV = 300 V improved the result to (45.3 $\pm$ 1.6) ps.

%The trends of the efficiency and time resolution as a function of the HV measured at this testbeam suggest that the nominal working point of the sensor was not reached yet. 
%{\color{red}At the time of the testbeam, it was not possible to increase beyond 300 V the HV of the board irradiated to \maxflu. This limitation was attributed (and verified after the testbeam) to radiation damage of the decoupling capacitors on the board hosting the ASIC.}
%In spite of this limitation, the partial 
The
scan in sensor bias voltage 
%possible at the testbeam for
of 
the board exposed to \maxflu, showed that the  plateau in detection efficiency was just reached at 300 V, while the time resolution did not yet reach its plateau value. These observations, supported by an estimation of the change in resistivity after \maxflu, imply that the sensor was not yet fully depleted at 300 V and suggest that the nominal working point at that proton fluence is  beyond 300 V.

%%It is worth mentioning that these 
%These timing results were obtained 
%without any dedicated radiation-tolerant design of the frontend electronics and 
%    with a monolithic sensor without  internal gain layer.
%Since the depleted region is 50 $\mu$m,  the detector can be thinned down to approximately 70 $\mu$m, including the CMOS processing. 
%We also remark that this sensor can achieve full fill factor.
%Thus it
%%does not require to be inclined inside a tracker by an angle to have uniform efficiency,
%%which would increase 
%minimises the silicon sensor area, the electronics channels and the services, with the consequence of reducing inactive material and the cost of a charged-particle tracking system. 

In conclusion, this study gives first hints that an amplifier using SiGe HBT is not affected greatly by doses up to \maxflu. The data show that very high efficiency, very high time resolution and radiation tolerance can be obtained by monolithic pixelated silicon detectors without internal gain layer that exploit low noise, low power and fast response SiGe HBTs. 
Commercial SiGe BiCMOS foundry processes therefore represent a very strong candidate to build monolithic silicon detectors for 4D tracking systems at future colliders at a contained cost.

\acknowledgments
This research is supported by the Horizon 2020 MONOLITH  ERC Advanced Grant ID: 884447. The authors wish to thank Coralie Husi, Javier Mesa, Gabriel Pelleriti, and the technical staff of the University of Geneva and IHP Microelectronics.
The authors acknowledge the support of EUROPRACTICE in providing design tools and MPW fabrication services, as well as the CERN SPS testbeam team.

\newpage
\bibliographystyle{unsrt}
\bibliography{bibliography.bib}
\end{document}